\documentclass[10pt,twocolumn,letterpaper]{article}

\usepackage{wacv}
\usepackage{times}
\usepackage{epsfig}
\usepackage{graphicx}
\usepackage{amsmath}
\usepackage{amssymb}

\usepackage{xcolor,colortbl}
\usepackage{multirow}

\wacvfinalcopy

\def\wacvPaperID{****} % Enter the WACV Paper ID here

\ifwacvfinal
\def\assignedStartPage{1}
\fi

\ifwacvfinal
\usepackage[breaklinks=true,bookmarks=false]{hyperref}
\else
\usepackage[pagebackref=true,breaklinks=true,colorlinks,bookmarks=false]{hyperref}
\fi

\ifwacvfinal
\else
\fi

\DeclareMathOperator*{\argmin}{arg\,min}
\usepackage{subfiles}

\begin{document}

%%%%%%%%% TITLE
%\title{An Attack on Co-Occurrence Based GAN Detectors}
\title{Adversarial Attacks on Co-Occurrence Features for GAN Detection}

\author{Michael Goebel\\
UC Santa Barbara\\
Santa Barbara, California\\
{\tt\small mgoebel@ucsb.edu}
\and
B.S. Manjunath\\
UC Santa Barbara\\
Santa Barbara, California\\
{\tt\small manj@ucsb.edu}
}

\maketitle
%\thispagestyle{empty}

%%%%%%%%% ABSTRACT
\begin{abstract}
Improvements in Generative Adversarial Networks (GANs) have greatly reduced the difficulty of producing new, photo-realistic images with unique semantic meaning. With this rise in ability to generate fake images comes demand to detect them. While numerous methods have been developed for this task, the majority of them remain vulnerable to adversarial attacks. In this paper, develop two novel adversarial attacks on co-occurrence based GAN detectors. These are the first attacks to be presented against such a detector. We show that our method can reduce accuracy from over 98\% to less than 4\%, with no knowledge of the deep learning model or weights. Furthermore, accuracy can be reduced to 0\% with full knowledge of the deep learning model details.

\end{abstract}

%%%%%%%%% BODY TEXT
\section{Introduction}

Since the advent of Generative Adversarial Networks (GANs) in 2014 \cite{goodfellow2014generative}, there has been a dramatic increase in the capabilities of GAN networks. They are now able to modify facial attributes in photo-realistic 1 megapixel images \cite{karras2019style}, perform image-to-image translation \cite{zhu2017unpaired}, and create new images given only a segmentation map \cite{park2019SPADE}.

With the vast increase in capability of GANs, there have been a number of detectors developed to distinguish GAN from authentic images \cite{nataraj2019detecting,wang2020cnn,zhang2019detecting,barni2020cnn,8397040}. These detectors generally perform well, reporting over 98\% test set accuracy. However, the question then arises as to what adversarial methods can be used to fool these detectors.

We focus primarily on attacking the co-occurrence feature based GAN detector proposed by Nataraj \etal \cite{nataraj2019detecting}. Our paper makes the following contributions:

\begin{enumerate}
    \item Describes the first targeted attacks against co-occurrence based GAN detectors known to the authors. 
    \item Generalizes the discrete co-occurrence matrices used for GAN detection to a differentiable function, and shows that gradient descent based attacks on a co-occurrence based detector with known model weights can be used to bring detector accuracy down to 0\%.
    \item Creates a new attack gray-box against co-occurrence based detectors, which assumes no knowledge of the deep learning weights or architecture. Tests with this method show that it can drop accuracy on GAN images from 98\% to less than 4\%.
    \item Demonstrate that without adversarial training against our method, it can decrease accuracy of other GAN detection methods by approximately 18\%.
%    \item Show that our gray-box method can also reduce accuracy on other detection methods without adversarial retraining, but will leave artifacts which can be detected after retaining of other detection methods.
\end{enumerate}

\begin{figure}
    \centering
    \includegraphics[width=0.95\linewidth]{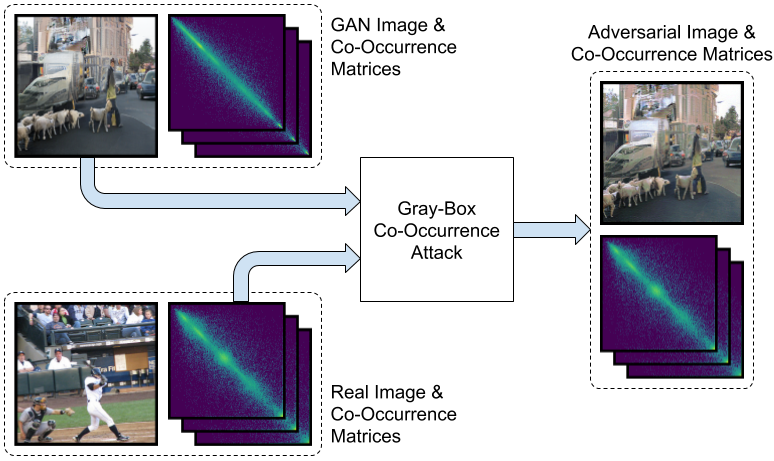}
    \caption{Outline of this paper's primary contribution, the gray-box co-occurrence attack. The method can impart a co-occurrence matrix taken from a benign image into a GAN image, fooling a GAN detector. Shown in each dashed box are the images and associated co-occurrence matrices.}
    \label{fig:high_level_bock_diagram}
\end{figure}

\section{Related Works}
\label{sec:related_works}

\subsection{Co-Occurrence Matrices in Steganography}

Co-occurrence matrices have a long history in image forensics, as features for detecting steganography, along with works intended to counter such detectors \cite{sullivan2005steganalysis,sullivan2006steganalysis}. Though steganography presents a unique challenge in that an adversarial image still must contain the embedded message, changes are often to a small number of pixels, and to a small degree, making restoring the statistics an easier task. In contrast, GAN manipulations are more pronounced, significantly changing the image's pixels and semantics.

%Works have been produced to counter these stego detectors, aiming to minimize perturbation to an image co-occurrence matrix after embedding a message. Though steganography presents a unique challenge in that an adversarial image still must contain the embedded message, changes are often to a small number of pixel, and to a minima degree, making restoring the statistics an easier task. In contrast, GAN images should be visually different than any real image at the adversaries disposal.

%Several works have been produced to counter co-occurrence based steganography detectors, aiming to make the tampered image have the same statistics as a real one. However, the challenges in this task are unique. For a real image and message, the adversary must change the image such that the message can be extracted, while minimally perturbing the image and co-occurrence matrix. For our task, our GAN and real image pairs are semantically and visually very different, and we wish to satisfy the same small difference. However, we are free of the steganography constraint of the LSBs carrying any significance.

\subsection{GAN Detection Methods}

Many methods of GAN detection have been proposed, though the one of primary interest will be that used in Nataraj \etal. They propose using co-occurrence matrices as a feature, which will be passed to a binary deep learning classifier. Another variation of the co-occurrence detector was recently posted on arXiv, which includes cross-channel co-occurrence pairs in the feature \cite{barni2020cnn}. Abridged results generalizing our attack to this method are in section \ref{sec:cband}.

%Though this GAN detection task is new, co-occurrence matrices have a long history in image forensics, as features for detecting steganography. Several works have produced to counter co-occurrence steganography detectors, aiming to make the tampered image have the same statistics as a real one. However, the challenges in this task are unique. For a real image and message, the adversary must change the image such that the message can be extracted, while minimally perturbing the image and ts co-occurrence matrix. For our task, our GAN and real image pairs are semantically and visually very different, and we wish to satisfy the same small difference. However, we are free of the steganography constraint of the LSBs carrying any significance.

Several methods were proposed in an older work from F. Marra \etal \cite{8397040}. These included constraining the input layer of a neural network classifier to a high-pass filter, computing co-occurrence matrices on high-pass filtered images, and reusing the original GAN discriminator. A more recent method proposed that simply training a ResNet model directly on a real vs GAN dataset was effective on most modern GANs \cite{wang2020cnn}.

Another work from Zhang \etal used the Discrete Fourier Transform (DFT) of each image channel as a feature to pass to a ResNet model \cite{zhang2019detecting}. This method took advantage of the upsampling artifacts created by most GANs.

%artifacts left behind from the upsampling in most GAN models.

%In section \ref{sec:detection_methods}, we describe the co-occurrence, DFT, and direct methods in detail.

\subsection{Adversarial Methods Against GAN Detectors}

In creating adversarial methods against GAN detectors, there is a spectrum of assumed knowledge levels about the system. At one extreme, everything about the detection system is known, including model weights. This scenario, referred to as white-box attack, has been investigated heavily in general computer vision tasks \cite{goodfellow2014explaining,kurakin2016adversarial,madry2017towards,carlini2017towards,szegedy2013intriguing,dong2018boosting,sabour2015adversarial,xiao2018spatially}. Most of these attacks use some variation of gradient descent through the network to perturb an input, such that the network gives the incorrect label. While such attacks require that the entire model be known and differentiable, they can generally achieve high degradation in accuracy for a small change in input. Prior to our work, direct and DFT based methods fit this differentiability criteria, but co-occurrence detectors did not.

At the other extreme are black-box methods, which attempt to degrade detector performance with no knowledge of the detection method being used. Though generally benign in practice, JPEG compression, blurring, and downsampling have all be investigated for their effects on detector performance. An interesting alternative in this category was proposed by Neves et al \cite{neves2019ganprintr}. They show that their method can decrease detection accuracy from greater than 99\% to 82-95\% on an Xception based detector, assuming no adversarial retraining. For the co-occurrence based detector, accuracy decreased by 14.7\%.

Between these two cases are gray-box attacks. One of these gray-box methods was proposed in \cite{zhang2019detecting}. It utilized the fact that most GANs have a similar architecture for upsampling. Their AutoGAN architecture would take in a real image, and attempt to reproduce this image at the output. However, the model would be constrained to use the same upsampling layers as were used in GANs, and would ideally introduce the same artifacts. The authors proposed this as an efficient way to generate more GAN-like images for detector training.

%The authors proposed the major benefit of this method as being a simplification of data collection for GAN detector training, as many GAN-like images could be generated from a large number of unlabeled natural images.

While the previous gray-box method can impart GAN artifacts on a real image, it cannot do the reverse. For most adversarial scenarios, the attacker would be interested in methods to make a GAN image appear real. The majority of this paper will focus on this direction, with the reverse case investigated in section \ref{sec:reverse_co}.

%though we demonstrate in section \ref{sec:reverse_co} that our algorithm also works in the reverse direction.

\section{Detection Methods}

\label{sec:detection_methods}

\begin{figure}
    \centering
    \includegraphics[width=0.95\linewidth]{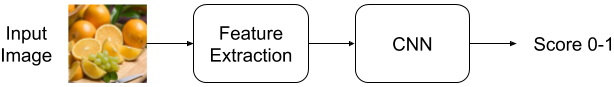}
    \caption{High-level diagram of the detection architectures. Sections \ref{sec:detection_co}, \ref{sec:detection_dft}, and \ref{sec:detection_direct} describe the different feature extraction methods investigated. Section \ref{sec:neural_net_selection} describes the different CNN architectures tested.}
    \label{fig:high_level_diagram}
\end{figure}

In addition to the co-occurrence features, we also investigate the co-occurrence and direct methods. Figure \ref{fig:high_level_diagram} shows an abstract outline of these models. This section describes the 3 methods we investigate for the feature extraction step.

%The primary focus of this paper will be on the co-occurrence based detection method, and adversarial attacks against it. However, as a point of comparison, we will also demonstrate results on two other detection methods.

\subsection{Co-Occurrence}
\label{sec:detection_co}

Unless stated otherwise, we will use the horizontal co-occurrence matrix defined by Nataraj \etal. For each channel in the RGB input image, represented by the array $X$, we produce a 2D histogram of horizontal pixel pairs:

\begin{equation} \label{eq:co_occurrence}
    C_{i,j} = \sum_{k,l} \delta(X_{k,l}-i) \cdot \delta(X_{k,l+1}-j)
\end{equation}

Where $\delta(\cdot)$ is the Kronecker delta function:

\begin{equation}
    \delta(k) =
    \begin{cases}
    1, \text{ if } k = 0 \\
    0, \text{ otherwise}
    \end{cases}, k \in \mathbb{Z}
\end{equation}

Each co-occurrence matrix is then scaled into the range $[0,1]$, re-stacked in the channel dimension, and passed to a deep learning classifier.

\subsection{DFT}
\label{sec:detection_dft}

As proposed by Zhang \etal, we will use the DFT of the image as an input to a deep learning classifier. The processing steps for the DFT based method are as follows:

\begin{enumerate}
    \item Get the centered, unitary DFT of the input image
    \item Take the magnitude of the DFT
    \item Apply the function $f(x) = log \left (x + 10^{-6} \right )$
    \item Shift and scale into the range [-1,1]
\end{enumerate}

All steps are consistent with those used in X. Zhang \etal, except for the inclusion of a small constant in step 3, to prevent a $log(0)$ case. They use a ResNet34 architecture pretrained with ImageNet weights for detection.

\subsection{Direct}
\label{sec:detection_direct}

This method will pass the image directly to a deep learning classifier, with only affine scaling as a preprocessing step. 
The following ImageNet means and standard are applied, as per the torchvision documentation \cite{marcel2010torchvision}:

%are applied as the default torchvision normalization \cite{marcel2010torchvision}:

%The following means and standard deviations are the default values in torchvision for each RGB channel, and derived from ImageNet \cite{marcel2010torchvision}:

\begin{equation}
    mean = 255 \cdot [0.485, 0.456, 0.406],
\end{equation}
\begin{equation}
    std = 255 \cdot [0.229, 0.224, 0.225]
\end{equation}

For their direct method, Wang \etal used ResNet50 pretrained on ImageNet \cite{wang2020cnn}.

\section{Co-Occurrence Gray-Box Attack}

\label{sec:co_occurrence_attack}
\subsection{Attack Formulation}

For our gray-box attack, we assume that it is known that co-occurrence matrices are the only feature used for detection, but we have no knowledge about the deep learning model used on these matrices. We also assume that the adversary has an arbitrary set of real images at their disposal. The goal of the adversary will be to modify each GAN image by some small amount, such that the co-occurrence matrix of the adversarial image is close to, if not exactly equal to, the co-occurrence matrix of a real image. 

%If the method is successful, and a GAN image can be made to have a co-occurrence matrix exactly equal to that of a real image, then co-occurrence based methods will be unable to differentiate real from GAN images.

The adversarial, real, and GAN images will be represented by $X_A$, $X_R$, and $X_G$ respectively. $F(\cdot)$ is the co-occurrence function, $Loss_1(\cdot,\cdot)$ and $Loss_2(\cdot,\cdot)$ are the loss functions to be defined later, and $\lambda$ is a user-defined constant. The adversarial image will be proposed as:

\begin{equation} \label{eq:optimizer}
    X_A = \argmin_{\tilde{X}} Loss_1(F(\tilde{X}),F(X_R)) + \lambda Loss_2(\tilde{X},X_G)
\end{equation}

\subsection{Distinctions Between Co-Occurrence and Histograms}
\label{sub_sec:cc_vs_hist}

Given the formulation in \ref{eq:optimizer}, it is worth noting that the Earth Mover's Distance (EMD) solves a similar optimization problem, and that there exist efficient approximations \cite{rubner1998metric}. For a pair of 2 dimensional histograms and for some pre-defined cost function between bins, this would produce a minimal cost transformation from one histogram to the other. However, non-edge pixels will appear twice in the co-occurrence histogram, once as the left pixel in the pair, and again as the right. Therefore, entries in the co-occurrence matrix cannot be individually manipulated to achieve this optimal transport, without inadvertently changing values at another location in the histogram.

To handle this entanglement between pairs, we use gradient descent to find an approximate minima.
%With gradient descent, the input to the optimization problem can directly be the image. The mapping from image to pixel pairs poses no problem for differentiation. 
However, this will require the functions $F$, $Loss_1$, and $Loss_2$ to be differentiable. The original definition of the co-occurrence function was over only integer inputs, posing a problem for the differentiability requirement. The next few sections break down the details of how $F$, $Loss_1$, and $Loss_2$ are selected.

%An alternative is proposed in section \ref{sec:diff_co_occur}. We also demonstrate the need for a non-trival loss function for $Loss_1$ in section \ref{sec:cc_loss}.

\subsection{Differentiable Extension of Co-Occurrence Function}
\label{sec:diff_co_occur}

In creating a differentiable extension of the co-occurrence matrix, we impose the following requirements:

\begin{enumerate}
    \item For integer inputs, $F(\cdot)$ must be equivalent to the original co-occurrence function.
    \item The sum of the histogram bins should equal the number of input elements.% Equivalently, the total contribution of each input element should always be 1.
    \item For all input elements, the contribution to each bin must be non-increasing with respect to distance from that bin.
    \item It must be differentiable over $\mathbb{R}_{[0,255]}^2$.
\end{enumerate}

Given the original co-occurrence formulation in equation \ref{eq:co_occurrence}, a simple extension would be to define a new one-dimensional function $f(\cdot)$ which will interpolate the delta function's integer values:

\begin{equation} \label{eq:diff_co_occurrence}
    C_{i,j} = \sum_{k,l} f(X_{k,l}-i) \cdot f(X_{k,l+1}-j)
\end{equation}

%Using this formulation, the gradients can easily be computed for all pixels. Below is the result for non-edge pixels:

%\begin{equation} \label{eq:deriv_of_cc}
%    \frac{\partial C_{i,j}}{\partial X_{m,n}} =
%    f'(X_{m,n}-j) \left ( f(X_{m,n-1}-i) + f(X_{m,n+1}-j) \right )
%\end{equation}

%When the above definition is used, the previous 4 requirements can be simplified into the following constraints on $f(\cdot)$:

Equation \ref{eq:diff_co_occurrence} consists only of additions, multiplications, and $f(\cdot)$, making gradient calculation straightforward. From this equation, the previous 4 requirements can be simplified into these requirements on $f(\cdot)$:

\begin{enumerate}
    \item $f(x) = \delta(x), \: x \in \mathbb{Z}$
    \item $f(x) = 1 - f(x-1), \: \forall x \in \mathbb{R}_{[0,1]}$
    \item $\frac{df}{dx} \leq 0$ for $x>0$, and $\frac{df}{dx} \geq 0$ for $x<0$
    \item $\frac{df}{dx}$ should be defined for all $x \in \mathbb{R}_{[-255,255]}$
\end{enumerate}

The combination of constraints 1 and 3 will require that $f(x) = 0$ for $x \notin (-1,1)$. Therefore, each pixel pair will contribute to at most 4 bins. This fact was taken advantage of in implementation, as opposed to computing the entire summation in equation \ref{eq:diff_co_occurrence}. Both the triangle and raise cosine shown below were tested as interpolation functions:

\begin{equation}
    tri(x) = 
    \begin{cases}
    1 - |x|, \text{ if } |x| < 1\\
    0, \text{ otherwise}
    \end{cases}
\end{equation}

\begin{equation}
    raised\_cos(x) = 
    \begin{cases}
    \frac{1 + cos(\pi x)}{2}, \text{ if } |x| < 1\\
    0, \text{ otherwise}
    \end{cases}
\end{equation}

For the triangle function, derivatives at $x=-1,0,1$ are undefined, so the average of the left and right derivatives is used. Raise cosine gave better results experimentally, and will be used for the remained of the tests.

%Much like the rectified linear unit (ReLU), the derivative of the triangle function is undefined for a small number of points. For these cases the average of the left and right derivatives is used. 

%The derivatives of the triangle function at 

%The raised cosine gave better experimental results and will be used for the remainder of the tests. 

\subsection{Co-Occurrence Loss Function}
\label{sec:cc_loss}

In this section, we provide intuitive reasoning and experimental justification for our selection of $Loss_1$. We provide several motivating examples for the 1D and 2D histogram cases. 
%For the histogram examples, we will provide lists of 1 or 2 dimensional values as a proxy for the list of image pixel pairs.
"Source" will correspond to the GAN input, "target" to the real input, and "solution" to the adversarial solution. For all of these examples, we will assume $\lambda=0$.

%Note that while the Earth Mover's Distance (EMD) can find the optimal solution to these histogram 

\subsubsection{One-Dimensional Example}
%To motivate our selection for the co-occurrence loss function, we will give several examples for the simpler 1D histogram case. For all of these examples, $\lambda$ will be set to 0, and $X_A$ initialized to $X_G$.

%Note that the earth mover's distance (EMD) can find an optimal solution to this problem in O(n) time, but only in the one-dimensional case. When moving to the 2 dimensional case, with interdependence between pairs, gradient descent seems to be a reasonable solution. However, examining a gradient descent method in the 1 dimensional case will provide simpler intuition.

Consider the case with a source of $[1,2,3]$, a target of $[2,3,4]$, and $\lambda = 0$. We would now like gradient descent to push the source towards the target, making their histograms equal. In this example, all derivatives should be negative.

%find a transformation on the source which will make the histograms of the transformed vector and the target vector equal. In the context of gradient descent, where only information about partial derivatives at the current point are available, we would like all derivatives at initialization to be negative in this example.

%For example $[1,2,3] \rightarrow [2,3,4]$, $[1,2,3] \rightarrow [4,2,3]$, or $[1,2,3] \rightarrow [3,2,3]$ would all be suitable. Regardless, we need the mass of points in the source to shift to the right. In the context of gradient descent, where only information about partial derivatives at the current point are available, we would like all derivatives at initialization to be negative in this example.

\begin{figure}
    \centering
    \includegraphics[width=0.95\linewidth]{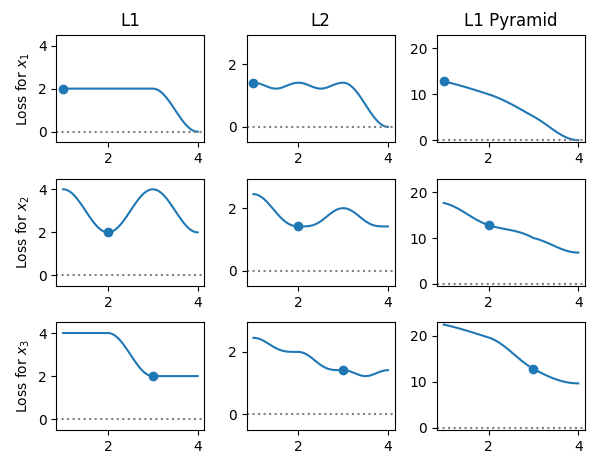}
    \caption{Loss functions with respect to value in each index for 1D loss functions, on source of [1,2,3] and target of [2,3,4]. Ideally, we would like the gradients for each index, at initialization, to be negative. This will push each value in the source towards the target. L1 loss would often get stuck on the plataeus, L2 would get stuck at the minima in-between integer values, and the L1 pyramid converged more easily. See figure \ref{fig:example_1d} for convergence results.}    
    \label{fig:univariate_loss_functions}
\end{figure}

Shown in Figure \ref{fig:univariate_loss_functions} are plots of loss for different loss functions, varying one input at a time. Consider the loss values as $x_1$ is moved from 1 to 4. For L1, there is a constant loss from $x_1 = 1$ to $3$. This is compared to L2, where loss fluctuates in the same region. Given that the histogram will have a constant L1 norm, and that L2 loss is less for the vector [1/2,1/2] than [0,1], the L2 loss function tended to get stuck between integer values. For this reason, we will focus our attention on L1 loss.

%favor converging between integer values. For the remainder of tests, we will consider only L1 based losses.

Looking at the top left graph in Figure \ref{fig:univariate_loss_functions}, the loss for $x_1$ decreases only after passing the threshold of 3. %In the current form, using point-wise loss and an interpolation function with a support of (-1,1), the vacancy at 4 can only pull values in which are within the (-1,1) support region.
Using point-wise loss, the vacancy at 4 can only pull values which are within the $(-1,1)$ support region of $f(\cdot)$.
To alleviate this, we instead compute loss on a multiscale pyramid of the histograms, with a downsampling factor of 2 in each step.
%To alleviate this, we instead compute loss on an image pyramid, with a downsampling factor of 2 in each step. 

%Consider the case of $X_G = [0,1,2]$ and $X_R = [1,2,3]$. For this problem, there are many solutions. The first value of 0 could be pushed all the way to 3, or 1 could be added to each value. Regardless, the histogram mass of $X_A$ needs to be shifted in the positive direction. Given only the ability to independently consider each variable at each step, the algorithm should give negative gradients at the initialization point. Shown in Figure \ref{fig:univariate_loss_functions}, simple L1 and L2 distance metrics fail to produce consistent negative gradients for each of the indices.

%Consider the loss values as $x_1$ is moved from 0 to 3. For L1, there is a constant loss from $x_1 = 0$ to $2$. This is compared to L2, where loss fluctuates in the same region. Given that the histogram will have a constant L1 norm, and that L2 loss is less for the vector [1/2,1/2] than [0,1], the L2 loss function was a less intuitive and worse performing choice empirically. 

%From now on, the focus will be placed on L1 loss. Looking at the top left graph in Figure \ref{fig:univariate_loss_functions}, the loss for $x_1$ decreases only after passing the threshold of 2. This is confirmed by equation \ref{eq:deriv_of_cc}, where the derivative only depends on its nearest neighbors. In the current form, using point-wise loss and an interpolation function with a support of (-1,1), the vacancy at 3 can only pull values in which are within the (-1,1) support region. To alleviate this, we instead compute loss on an image pyramid, with a downsampling factor of 2 in each step. 

\begin{figure}
    \centering
    \includegraphics[width=0.95\linewidth]{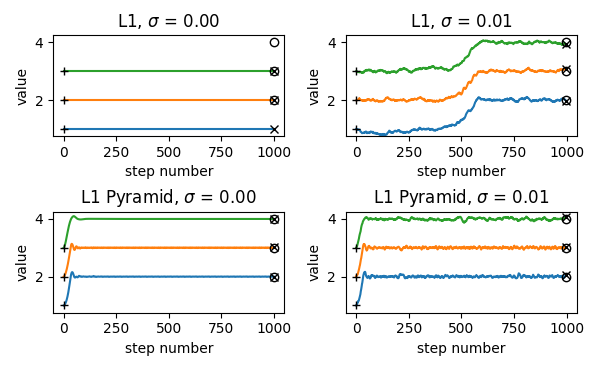}
    \caption{Plot of each value in the source vector with respect to step number, when solving with a source of [1,2,3] and target of [2,3,4].
    %This provides further evidence for the insuffieciency of simple $L_1$ for this task.
    Top left: without random noise, all points are stuck in flat regions, no change in values. Top right: the addition of noise allows for the algorithm using $L_1$ norm to gradually drift towards the target. Bottom row: Both with and without noise, the algorithm converges over 14 times faster than the top right case.}
    \label{fig:example_1d}
\end{figure}

To combine the multi-scale losses, a simple weighted sum is used. Weights are set equal to the downsampling factor at each level. This weight selection is motivated by the fact that the cost of moving pixels between bins at each layer in the pyramid scales with respect to the downsampling factor. Results using the image pyramid loss are shown in the right column of Figure \ref{fig:univariate_loss_functions}. When gradient descent is run on the different loss functions for the 1D case, the results in Figure \ref{fig:example_1d} are produced. 

%To combine these multi-scale losses into a single value, a each loss is weighted by its associated downsampling factor and 

%into a single value, a weighted summation is used, with the weights equal to the downsampling factor. Results on the toy problem are shown in the right two columns of Figure \ref{fig:univariate_loss_functions}.

\subsubsection{Two-Dimensional Examples}

\begin{figure}
    \centering
    \includegraphics[width=1.0\linewidth]{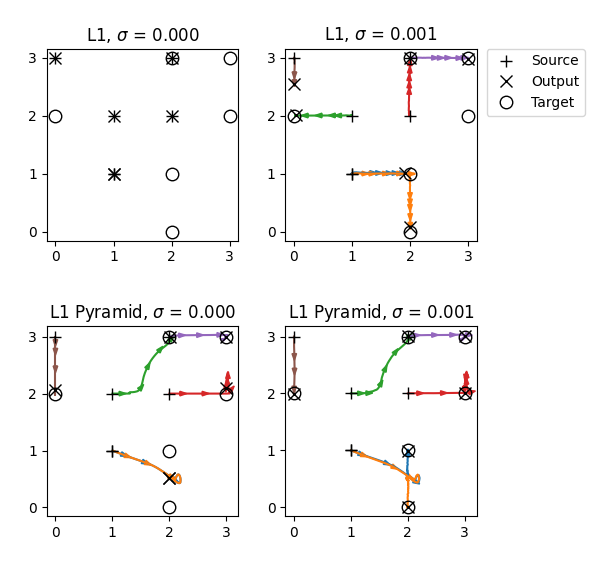}
    \caption{Example applying point-wise and pyramid loss to a 2D histogram gradient descent problem. Horizontal and vertical axes represent location of the 2D vectors. Top left: with $L_1$ loss and no noise, all gradients are 0. Top right: With noise, $L_1$ finds a sub-optimal minima, where not all target points are reached.
    %However, it gets stuck in a sub-optimal solution. The green path (originating from (1,2)) should have gone right, to fill the remaining empty target point.
    Bottom left: Without added noise, the pyramid loss almost converges to a global minima. However, the two source points originating from (1,1) need to split and fill different target points. With deterministic gradient descent, this splitting will not occur.
    %However, two source points originate at (1,1), and need to split to fill different points in the target. With deterministic gradient descent, this splitting cannot happen.
    Bottom right: A proper solution is found with pyramid loss and  noise.}
    \label{fig:example_2d}
\end{figure}

A 2D example is shown in figure \ref{fig:example_2d}. Especially important from this figure is the necessity of random noise. Often we will need points initialized to the same value to arbitrarilly split into two different outcomes, which cannot occur with deterministic gradient descent.

%With regular gradient descent, no arbitrary splitting occurs, and both values get stuck between the two target points.

This two-dimensional histogram test was repeated over 100 iterations, with both source and target vectors containing 8 elements uniformally sampled from $\mathbb{Z}_{[0,7]}^2$. For the L1 pyramid with Gaussian noise, all 100 tests successfully converged from the source to target. This is compared to only 8 for L1 with Gaussian noise.

\subsubsection{Extension to Co-Occurrence on Images}

For an 8-bit image, a 9 layer pyramid is used, with downsampling factors from 1 (None) to 256. To implement the blurring and downsampling steps in the co-occurrence image pyramid, we rely upon the original interpolation function defined for the co-occurrence. By dividing the input image by the downsampling factor before computing co-occurrence, lower resolution co-occurrence matrices can be produced. The full loss function is shown in equation \ref{eq:pyramid_loss}.
%Again, $F(\cdot)$ is the differentiable co-occurrence function defined in equation \ref{eq:co_occurrence}.

\begin{equation} \label{eq:pyramid_loss}
    loss = \sum_{n=0}^{8} 2^{n} \left \| F \left (\frac{X_A}{2^n} \right ) - F \left (\frac{X_R}{2^n} \right ) \right \|_{1}
\end{equation}

\subsection{Image-Space Loss Function}

In equation \ref{eq:optimizer}, only the $Loss_2$ and $\lambda$ terms are left to be defined. For consistency with the co-occurrence loss, L1 distance is chosen for the image-space loss. The $\lambda$ parameter remains as a user selected parameter, and several values were tested experimentally. 

\subsection{Implementation Details}

%With many possible GAN-real image pairs, we would like to choose pairs with similar color values before optimization. 
Ideally, we would like to choose source-target pairs with similar color values for optimization. For example, we would not want to force a GAN image with a green grassy background to have the same color distribution as a real image of a blue ocean. To do this, we divide the data into blocks of size approximately 900, and for each GAN image, select the real image whose EMD over the 1D RGB histograms is closest to that of the GAN image.

With the pairs selected, we can then run our gradient descent algorithm. The solution is initialized with the source image. We use a standard gradient descent, with a learning rate of 0.01, and momentum of 0.9. This is done in 3 sequential epochs, with 200, 50, and 50 steps. For the first 2 epochs, Gaussian noise with standard deviation of 0.01 is added to the image. No noise is added in the last epoch. The solution is rounded after each epoch.

%At the end of each block, the image is also rounded to the nearest integer.

When run on an Nvidia 1080 Ti, the algorithm took approximately 30 seconds per 256x256 image. However, up to 3 processes could be placed on a single GPU, so 6 adversarial images could be produced every minute. 

\begin{figure}
    \centering
    \includegraphics[width=0.95\linewidth]{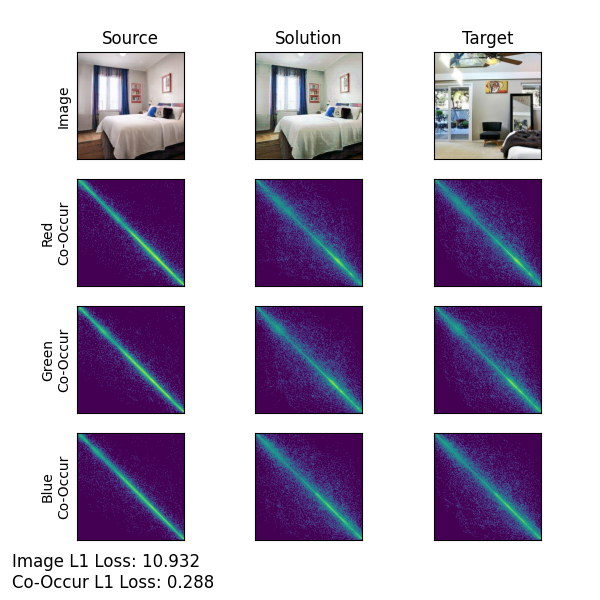}
    \caption{Example solution found by our algorithm, for the given real and GAN images.}
    \label{fig:good_solution_example}
\end{figure}

\begin{figure}
    \centering
    \includegraphics[width=0.95\linewidth]{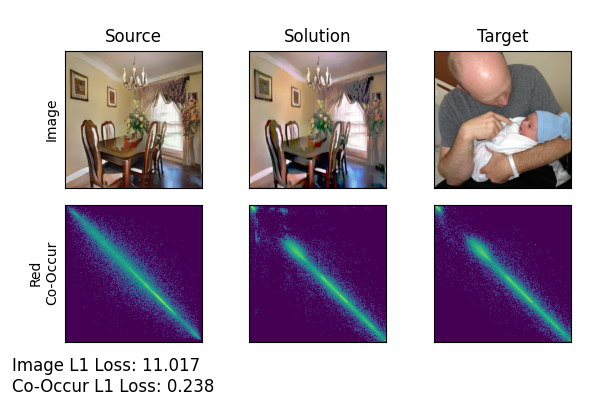}
    \caption{Another example solution found by our algorithm, and corresponding red channel co-occurrence matrices. In the top-left corner of the solution co-occurrence a square artifact can be seen.}
    \label{fig:bad_solution_example}
\end{figure}

\begin{figure}
    \centering
    \includegraphics[width=0.95\linewidth]{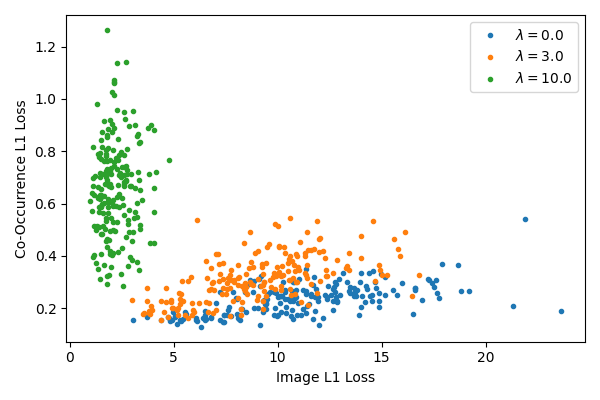}
    \caption{Testing of the co-occurrence algorithm on our dataset described in section \ref{sec:datasets} on 200 images for 3 different $\lambda$ values. Smaller $\lambda$ values will force the co-occurrence matrix of the adversarial image closer to that of the real image, at the cost of greater perturbation.}
    \label{fig:cc_loss_scatter}
\end{figure}

Quantitative results are shown in figure \ref{fig:cc_loss_scatter}. For comparison, average $L_1$ loss between the co-occurrence matrices of the source target pairs was 0.90, and $L_1$ loss between source and target images was 52.7. For this real data, it cannot achieve a perfect match between the real and adversarial co-occurrence matrices. Two examples are given in figures \ref{fig:good_solution_example} and \ref{fig:bad_solution_example}. The effect of this slight mismatch is investigated experimentally.

\section{Other Adversarial Attacks}
\subsection{Gray-Box DFT}

This method follows a similar formulation to the co-occurrence gray-box attack. A real image is obtained in addition to the GAN image, and the detection feature of the adversarial image is made to be similar to the real image, while minimizing the distance from the original GAN image. We rely upon the intuition that the defining features of the GAN in the DFT domain are concentrated away from the DC axes. To estimate this high-frequency noise signal, we use the same filter as Kirchner in his work on resampling \cite{kirchner2008fast}, and has a centered DFT given in equation \ref{eq:kirchner_dft}:

\begin{equation}
    \label{eq:kirchner_dft}
    \mathcal{F}(f) = \frac{1}{4}
\begin{bmatrix}
    3 & 0 & 3 \\
    0 & 0 & 0 \\
    3 & 0 & 3
\end{bmatrix}
\end{equation}

%\begin{equation}
%f = \frac{1}{4}
%\begin{bmatrix}
%    1 & -2 & 1 \\
%    -2 & 4 & -2 \\
%    1 & -2 & 1
%\end{bmatrix}
%\end{equation}

%Here, the origin is located in the middle of the 3x3 array. The unitary DFT is as follows, again with the origin at the center of the 3x3 array:

%\begin{equation}
%    \mathcal{F}(f) = \frac{1}{4}
%\begin{bmatrix}
%    3 & 0 & 3 \\
%    0 & 0 & 0 \\
%    3 & 0 & 3
%\end{bmatrix}
%\end{equation}

To produce an adversarial image, we solve the following:

\begin{equation} \label{eq:dft_optimization}
    X_A = \argmin_{\tilde{X}} \left \Vert f*\tilde{X} - f*X_R \right \Vert_{2}^{2} + \lambda^2 \left \Vert \tilde{X} - X_G \right \Vert_{2}^{2}
\end{equation}

Application of the Fourier transform turns this problem into one of weighted least squares, and can be solved as:

\begin{equation} \label{eq:dft_solution}
    \mathcal{F}(X_A) = \frac{ \mathcal{F}(f)^2 \cdot \mathcal{F}(X_R) + \lambda^2 \mathcal{F}(X_G) }{ \mathcal{F}(f)^2 + \lambda^2}
\end{equation}

This is done between randomly selected real and GAN images, for different $\lambda$ values in section \ref{sec:experiments}.

\subsection{White-Box PGD}

With the co-occurrence, DFT, and direct methods all being differentiable pytorch functions, we run the $L_{\infty}$ PGD algorithm on all GAN images on each method \cite{madry2017towards}. This is done using the advertorch library \cite{ding2019advertorch}. Default parameters are used, with a maximum distortion of 1, maximum step size of 2/40, and 40 total iterations, running on pixels in the range $[0,255]$. Pixels are rounded after completion of PGD.

%After running PGD, pixel values are rounded to the nearest integer.

%With the co-occurrence, DFT, and direct methods all being differentiable pytorch functions, we run the $L_{\infty}$ PGD algorithm on all GAN images on each method \cite{madry2017towards}. PGD is an iterative algorithm, which uses backpropagation through the network to find changes in the image which will increase the predicted score of a desired output. 

%This is done using the advertorch library \cite{ding2019advertorch}. Default parameters are used, with a maximum distortion of 1, maximum step size of 2/40, and 40 total iterations, running on pixels in the range $[0,255]$. After running PGD, pixel values are rounded to the nearest integer.

\section{Experiments}
\label{sec:experiments}

\subsection{Datasets}
\label{sec:datasets}

Our dataset consists of 4 different GANs, drawing from a variety of image datasets and tasks. Image counts are given in table \ref{tab:dataset_overview}. The each group is divided into a 70/15/15 train/val/test split. These groups are then further split in half; the first for training and testing of models, and the second for generating adversarial samples. All images were center-cropped to 256x256.

%The first half will be used for training and testing models, the other half will be used to generate adversarial images for their respective train, validation, or test group. For images larger than 256x256, we take a center crop.  

\begin{table}
\begin{center}
\scalebox{0.75}{
\begin{tabular}{|c|c|c|}
    \hline
    Architecture & Dataset & Total Count \\
    \hline \hline
    CycleGAN \cite{zhu2017unpaired} & apple2orange \cite{deng2009imagenet} & 3,000 \\ 
     & horse2zebra \cite{deng2009imagenet} & 3,000 \\
     & summer2winter \cite{zhu2017unpaired} & 3,000 \\
     & cityscapes \cite{cordts2016cityscapes} & 3,000 \\
     & cezanne \cite{zhu2017unpaired} & 3,000 \\ 
     & monet \cite{zhu2017unpaired} & 3,000 \\
     & ukiyoe \cite{zhu2017unpaired} & 3,000 \\
     & vangogh \cite{zhu2017unpaired} & 3,000 \\
    \hline
    ProGAN \cite{karras2017progressive} & CelebHQ \cite{liu2015faceattributes} & 24,000 \\
    \hline
    SPADE \cite{park2019SPADE} & ADE20K \cite{zhou2019semantic} & 12,000 \\
     & COCO-Stuff \cite{caesar2018coco} & 12,000 \\
    \hline
    StyleGAN \cite{karras2019style} & LSUN Bedroom \cite{yu2015lsun} & 8,000 \\
     & LSUN Car \cite{yu2015lsun} & 8,000 \\
     & LSUN Cat \cite{yu2015lsun} & 8,000 \\
    \hline
    \hline
    Total & & 96,000 \\
    \hline
     
\end{tabular}
}
\caption{Combined number of real and fake samples from each data subset. For all subsets, the number of real and fake examples is equal. In all, there were 48k real and 48k GAN images used.}
\label{tab:dataset_overview}
\end{center}
\end{table}

\subsection{Neural Network Selection}
\label{sec:neural_net_selection}

Neural network selection was done experimentally, and results are shown in table \ref{tab:diff_networks}. All models were trained with 16 real and 16 GAN images per batch, for 16 epochs. An Adam optimizer was used, with default parameters of 0.001 for the learning rate, and (0.9,0.999) for the betas \cite{kingma2014adam}. The final model weights for testing are selected from the epoch number on which validation loss was the lowest.

%The ImageNet pretrained networks did better than those with random initializations in almost all cases. Several of the larger networks failed to converge. Of the networks that did converge, the results remained relatively close for varying capacities. Given that there does not seem to be a benefit to larger network on this dataset, we will proceed using only pretrained ResNet18 and MobileNet. ResNet18 was selected as the model against which PGD samples would be created.

The ImageNet pretrained networks did better than those with random initializations in almost all cases. As the larger networks did not provide noticeable improvements on this dataset, we will use pretrained ResNet18 and pretrained MobileNet for the remainder of tests. We chose ResNet18 to generate the PGD samples, given that ResNets were used by both Zhang \etal and Wang \etal.

\begin{table}
    \centering
    \scalebox{0.75}{
\begin{tabular}{|c||c|c|c|c|c|c|c|c|c|}
\hline
Method & \multicolumn{2}{c|}{Co-Occurrence}  & \multicolumn{2}{c|}{DFT} & \multicolumn{2}{c|}{Direct} \\
\hline
Initialization & ImNet & rand & ImNet & rand & ImNet & rand \\
\hline
ResNet18 & 0.979 \cellcolor[rgb]{0.973,0.950,0.556} & 0.974 \cellcolor[rgb]{0.968,0.949,0.554} & 0.904 \cellcolor[rgb]{0.876,0.937,0.572} & 0.888 \cellcolor[rgb]{0.855,0.934,0.585} & 0.980 \cellcolor[rgb]{0.973,0.950,0.556} & 0.829 \cellcolor[rgb]{0.778,0.920,0.635} \\ 
ResNet50 & 0.979 \cellcolor[rgb]{0.973,0.950,0.556} & 0.977 \cellcolor[rgb]{0.973,0.950,0.556} & 0.864 \cellcolor[rgb]{0.824,0.929,0.605} & 0.900 \cellcolor[rgb]{0.871,0.937,0.575} & 0.976 \cellcolor[rgb]{0.968,0.949,0.554} & 0.866 \cellcolor[rgb]{0.824,0.929,0.605} \\ 
ResNet101 & 0.424 \cellcolor[rgb]{0.577,0.747,0.779} & 0.569 \cellcolor[rgb]{0.561,0.815,0.766} & 0.503 \cellcolor[rgb]{0.564,0.783,0.775} & 0.500 \cellcolor[rgb]{0.564,0.783,0.775} & 0.519 \cellcolor[rgb]{0.562,0.791,0.774} & 0.503 \cellcolor[rgb]{0.564,0.783,0.775} \\ 
ResNet152 & 0.500 \cellcolor[rgb]{0.564,0.783,0.775} & 0.668 \cellcolor[rgb]{0.607,0.861,0.735} & 0.495 \cellcolor[rgb]{0.565,0.780,0.776} & 0.500 \cellcolor[rgb]{0.564,0.783,0.775} & 0.500 \cellcolor[rgb]{0.564,0.782,0.776} & 0.499 \cellcolor[rgb]{0.564,0.782,0.776} \\ 
ResNeXt50 & 0.978 \cellcolor[rgb]{0.973,0.950,0.556} & 0.975 \cellcolor[rgb]{0.968,0.949,0.554} & 0.882 \cellcolor[rgb]{0.844,0.933,0.591} & 0.907 \cellcolor[rgb]{0.881,0.938,0.569} & 0.986 \cellcolor[rgb]{0.982,0.951,0.562} & 0.853 \cellcolor[rgb]{0.808,0.926,0.615} \\ 
Inception V3 & 0.944 \cellcolor[rgb]{0.928,0.944,0.549} & 0.500 \cellcolor[rgb]{0.564,0.783,0.775} & 0.948 \cellcolor[rgb]{0.933,0.945,0.548} & 0.708 \cellcolor[rgb]{0.641,0.878,0.716} & 0.990 \cellcolor[rgb]{0.987,0.952,0.565} & 0.949 \cellcolor[rgb]{0.933,0.945,0.548} \\ 
MobileNet & 0.978 \cellcolor[rgb]{0.973,0.950,0.556} & 0.974 \cellcolor[rgb]{0.968,0.949,0.554} & 0.949 \cellcolor[rgb]{0.938,0.946,0.548} & 0.919 \cellcolor[rgb]{0.897,0.940,0.560} & 0.996 \cellcolor[rgb]{0.997,0.953,0.572} & 0.989 \cellcolor[rgb]{0.987,0.952,0.565} \\ 
\hline
\end{tabular}

    }
    \caption{Overall accuracy of different networks on a balanced test set of real and GAN. Each row represents using a different deep learning architecture. Three detection methods are shown as the first column headers. The second shows results with either ImageNet weights or random initialization.}
    %and the second show the difference between initializing network weights randomly, or with pre-trained ImageNet weights. From this more comprehensive test, we decided to limit our adversarial retraining to only pretrained ResNet18 and pretrained MobileNet.}
    \label{tab:diff_networks}
\end{table}

\subsection{Testing on Adversarial Samples}

We then evaluated the detectors chosen in the previous section on the co-occurrence adversarial examples, with results shown in Table \ref{tab:results_no_retrain}. For both detectors, the gray-box co-occurrence attack drops the GAN detection rate from approximately 98\% to less than 4\%, with no knowledge of the deep-learning model used. As expected with the PGD attack, accuracy on the exact model being attacked drops to 0. However, accuracy on MobileNet drops to only 8\%; more than twice what was achieved with the gray-box attack.

\begin{table}
    \centering
    \scalebox{0.78}{
    \begin{tabular}{|c||c|c|c|c|}
        \hline
         & Real & GAN & GB CO $\lambda=0.0$ & PGD CO \\
         \hline \hline
         ResNet18 & 0.979 \cellcolor[rgb]{0.973,0.950,0.556} & 0.984 \cellcolor[rgb]{0.978,0.951,0.559} & 0.030 \cellcolor[rgb]{0.638,0.522,0.685} & 0.000 \cellcolor[rgb]{0.634,0.502,0.665} \\
         MobileNet & 0.976 \cellcolor[rgb]{0.968,0.949,0.554} & 0.981 \cellcolor[rgb]{0.978,0.951,0.559} & 0.039 \cellcolor[rgb]{0.639,0.531,0.693} & 0.083 \cellcolor[rgb]{0.642,0.560,0.720} \\
         \hline
    \end{tabular}
    }
    \caption{Test set accuracy of co-occurrence based detectors, without adversarial retraining. Gray-box (GB) co-occurrence (CO) samples are generated as described in section \ref{sec:co_occurrence_attack}. PGD co-occurrence examples are produced using ResNet18.}
    \label{tab:results_no_retrain}
\end{table}

\subsection{Adversarial Training}

Next we adversarially trained the same networks using different subsets of the adversarial samples. The labels remain binary, with real images in one class, and all GAN images, including adversarial GAN images, in the other.
%The effects of training on different subsets of adversarial examples is also investigated.

%We investigate the effects of training on multiple different subsets of all of the adversarial examples as well.

%For adversarial retraining, we need a more complex data balancing strategy than in used in the original training procedure.

For data balancing, we maintain an equal number of positive and negative samples. Within the positive sample class, each of the sub-types is sampled equally. For example, in the test using all gray-box co-occurrence adversarial examples, we used 16 real, 4 GAN, and 4 from each of the 3 gray-box co-occurrence classes.
%A batch size of 32 is used in all cases except for the set which includes all adversarial images. To evenly divide these images, we use a batch size of 40.
For the set of all adversarial images, a batch size of 40 is used so the batch can be evenly divided. For all other cases, batch size remains 32.

%\begin{itemize}
%    \item No Adv: 16 real, 16 GAN
%    \item GB CC $\lambda=0.0$: 16 real, 8 GAN, 8 gray-box CC $\lambda=0.0$
%    \item GB CC All: 16 real, 4 GAN, 4 from each of the 3 gray-box CC set
%    \item All Adv: 20 Real, 2 GAN, 2 from each of the 9 adversarial sets
%\end{itemize}

%\subsubsection{Co-Occurrence Detector}

\begin{table}[]
    \centering
    \scalebox{0.75}{

%\begin{tabular}{|c|c|c|c|c|}
%\hline
%    \multicolumn{2}{|c|}{} & \multicolumn{3}{c|}{Test} \\
%    \cline{3-5}
%    \multicolumn{2}{|c|}{} & Real & GAN & CO $\lambda=0$ \\
%    \hline
%    \multirow{2}{*}{Train} & \shortstack{Real \& \\ GAN} & & & \\
%    \cline{2-5}
%    & \shortstack{Real, GAN, \\ \& CO $\lambda=0$} & & & \\
%    \hline
%\end{tabular}

%\begin{tabular}{|c|c|c|c|c|}
%\hline
%    \multicolumn{2}{|c|}{} & \multicolumn{3}{c|}{Test} \\
%    \cline{3-5}
%    \multicolumn{2}{|c|}{} & Real & GAN & CO $\lambda=0$ \\
%    \hline
%    \multirow{2}{*}{Train} & Real, GAN &  0.979 \cellcolor[rgb]{0.973,0.950,0.556} & 0.984 \cellcolor[rgb]{0.978,0.951,0.559} & 0.030 \cellcolor[rgb]{0.638,0.522,0.685} \\
%    \cline{2-5}
%    & \shortstack{Real, GAN, \\ \& CO $\lambda=0$} & & & \\
%    \hline
%\end{tabular}

\begin{tabular}{|c|c|c|c|c|c|}
\hline
    \multicolumn{2}{|c|}{} & \multicolumn{4}{c|}{Test} \\
    \cline{3-6}
    \multicolumn{2}{|c|}{} & Real & GAN & CO $\lambda=0$ & CO PGD \\
    \hline
    \multirow{2}{*}{Train} & Real, GAN &  0.979 \cellcolor[rgb]{0.973,0.950,0.556} & 0.984 \cellcolor[rgb]{0.978,0.951,0.559} & 0.030 \cellcolor[rgb]{0.638,0.522,0.685} & 0.000 \cellcolor[rgb]{0.634,0.502,0.665} \\
    \cline{2-6}
    & Real, GAN, All Adv & 0.901 \cellcolor[rgb]{0.871,0.937,0.575} & 0.971 \cellcolor[rgb]{0.963,0.949,0.552} & 0.970 \cellcolor[rgb]{0.963,0.949,0.552} & 1.000 \cellcolor[rgb]{0.997,0.953,0.572} \\
    \hline
\end{tabular}

%\shortstack{Real, GAN,\\ All Adv}

    }
    \caption{Results for only the ResNet18 co-occurrence detector. The rows show results with and without adversarial retraining.}
    \label{tab:results_co_only}
\end{table}

%For the co-occurrence based detector trained on gray-box adversarial co-occurrence images, average accuracy was still slightly lower than the detector trained and tested without adversarial images. Interestingly, attacks targeted towards a particular method also worsen results on detectors based on a different methods. For the co-occurrence based detector in the first row, gray-box DFT, PGD DFT, and PGD direct images have accuracies of 0.62, 0.86, and 0.85. In most cases, these loses were recouped after adversarial training against these methods. 

%From these results, defense against an attack method almost always requires training on that particular method. For the entire gray-box co-occurrence column with $\lambda = 0$, detectors which were not trained on this set all got below 87\% accuracy, while those which were trained were all above 94\%. Though this method was specifically targeted at the co-occurrence matrix of the image, it produced at least a 10\% drop in accuracy on all other non-adversarially trained detection methods as well.

\subsubsection{Full Results}

\begin{table*}[ht]
\centering
\scalebox{0.7}{
\begin{tabular}{|c|c|c|c|c|c|c|c|c|c|c|c|c|c|c|c|c|c|c|c|c|c|c|c|c|c|c|c|c|c|c|}
\hline
 \multicolumn{3}{|c|}{} & Real & GAN & \multicolumn{3}{c|}{Co-Occur Gray-Box} & \multicolumn{3}{c|}{DFT Gray-Box} & \multicolumn{3}{c|}{PGD} \\
 \multicolumn{3}{|c|}{} & & & $\lambda = 0.0$ & $\lambda = 3.0$ & $\lambda = 10.0$ & $\lambda = 0.003$ & $\lambda = 0.01$ & $\lambda = 0.03$ & Co-Occur & DFT & Direct \\
\hline
\hline 

\multirow{21}{*}{ResNet18} & \multirow{9}{*}{Co-Occur} & No Adv* & 0.979 \cellcolor[rgb]{0.973,0.950,0.556} & 0.984 \cellcolor[rgb]{0.978,0.951,0.559} & 0.030 \cellcolor[rgb]{0.638,0.522,0.685} & 0.019 \cellcolor[rgb]{0.636,0.513,0.677} & 0.332 \cellcolor[rgb]{0.595,0.704,0.778} & 0.624 \cellcolor[rgb]{0.577,0.840,0.752} & 0.627 \cellcolor[rgb]{0.579,0.842,0.751} & 0.627 \cellcolor[rgb]{0.579,0.842,0.751} & 0.000 \cellcolor[rgb]{0.634,0.502,0.665} & 0.863 \cellcolor[rgb]{0.818,0.928,0.608} & 0.854 \cellcolor[rgb]{0.808,0.926,0.615} \\
 &  & No Adv & 0.976 \cellcolor[rgb]{0.968,0.949,0.554} & 0.980 \cellcolor[rgb]{0.973,0.950,0.556} & 0.032 \cellcolor[rgb]{0.639,0.525,0.688} & 0.027 \cellcolor[rgb]{0.637,0.519,0.682} & 0.358 \cellcolor[rgb]{0.590,0.715,0.779} & 0.906 \cellcolor[rgb]{0.876,0.937,0.572} & 0.883 \cellcolor[rgb]{0.850,0.934,0.588} & 0.852 \cellcolor[rgb]{0.808,0.926,0.615} & 0.000 \cellcolor[rgb]{0.634,0.502,0.665} & 0.907 \cellcolor[rgb]{0.881,0.938,0.569} & 0.878 \cellcolor[rgb]{0.839,0.932,0.595} \\
 &  & GBCO 0.0 & 0.972 \cellcolor[rgb]{0.963,0.949,0.552} & 0.966 \cellcolor[rgb]{0.958,0.948,0.550} & 0.957 \cellcolor[rgb]{0.943,0.946,0.548} & 0.992 \cellcolor[rgb]{0.987,0.952,0.565} & 0.932 \cellcolor[rgb]{0.912,0.942,0.553} & 0.473 \cellcolor[rgb]{0.568,0.771,0.777} & 0.480 \cellcolor[rgb]{0.568,0.772,0.777} & 0.476 \cellcolor[rgb]{0.568,0.771,0.777} & 0.144 \cellcolor[rgb]{0.638,0.597,0.748} & 0.822 \cellcolor[rgb]{0.768,0.918,0.641} & 0.794 \cellcolor[rgb]{0.734,0.909,0.662} \\
 &  & GB-CO 3.0 & 0.974 \cellcolor[rgb]{0.968,0.949,0.554} & 0.972 \cellcolor[rgb]{0.963,0.949,0.552} & 0.424 \cellcolor[rgb]{0.577,0.747,0.779} & 0.985 \cellcolor[rgb]{0.982,0.951,0.562} & 0.999 \cellcolor[rgb]{0.997,0.953,0.572} & 0.702 \cellcolor[rgb]{0.633,0.874,0.720} & 0.685 \cellcolor[rgb]{0.620,0.868,0.728} & 0.666 \cellcolor[rgb]{0.604,0.859,0.736} & 0.311 \cellcolor[rgb]{0.601,0.692,0.777} & 0.923 \cellcolor[rgb]{0.902,0.941,0.557} & 0.910 \cellcolor[rgb]{0.886,0.939,0.566} \\
 &  & GB-CO 10.0 & 0.985 \cellcolor[rgb]{0.982,0.951,0.562} & 0.973 \cellcolor[rgb]{0.968,0.949,0.554} & 0.121 \cellcolor[rgb]{0.640,0.583,0.738} & 0.826 \cellcolor[rgb]{0.773,0.919,0.638} & 0.997 \cellcolor[rgb]{0.997,0.953,0.572} & 0.360 \cellcolor[rgb]{0.590,0.717,0.779} & 0.402 \cellcolor[rgb]{0.582,0.736,0.779} & 0.444 \cellcolor[rgb]{0.574,0.756,0.779} & 0.000 \cellcolor[rgb]{0.634,0.502,0.665} & 0.890 \cellcolor[rgb]{0.855,0.934,0.585} & 0.876 \cellcolor[rgb]{0.839,0.932,0.595} \\
 &  & All GB-CO & 0.964 \cellcolor[rgb]{0.953,0.947,0.549} & 0.951 \cellcolor[rgb]{0.938,0.946,0.548} & 0.947 \cellcolor[rgb]{0.933,0.945,0.548} & 0.999 \cellcolor[rgb]{0.997,0.953,0.572} & 0.997 \cellcolor[rgb]{0.997,0.953,0.572} & 0.415 \cellcolor[rgb]{0.579,0.743,0.779} & 0.448 \cellcolor[rgb]{0.573,0.758,0.778} & 0.472 \cellcolor[rgb]{0.569,0.769,0.777} & 0.104 \cellcolor[rgb]{0.641,0.573,0.731} & 0.883 \cellcolor[rgb]{0.850,0.934,0.588} & 0.858 \cellcolor[rgb]{0.813,0.927,0.612} \\
 &  & All GB-DFT & 0.968 \cellcolor[rgb]{0.958,0.948,0.550} & 0.984 \cellcolor[rgb]{0.982,0.951,0.562} & 0.047 \cellcolor[rgb]{0.640,0.537,0.699} & 0.029 \cellcolor[rgb]{0.638,0.522,0.685} & 0.222 \cellcolor[rgb]{0.622,0.644,0.769} & 0.998 \cellcolor[rgb]{0.997,0.953,0.572} & 0.997 \cellcolor[rgb]{0.997,0.953,0.572} & 0.994 \cellcolor[rgb]{0.992,0.952,0.568} & 0.000 \cellcolor[rgb]{0.634,0.502,0.665} & 0.893 \cellcolor[rgb]{0.860,0.935,0.581} & 0.880 \cellcolor[rgb]{0.844,0.933,0.591} \\
 &  & All PGD & 0.974 \cellcolor[rgb]{0.968,0.949,0.554} & 0.974 \cellcolor[rgb]{0.968,0.949,0.554} & 0.031 \cellcolor[rgb]{0.638,0.522,0.685} & 0.045 \cellcolor[rgb]{0.640,0.534,0.696} & 0.449 \cellcolor[rgb]{0.573,0.758,0.778} & 0.832 \cellcolor[rgb]{0.783,0.921,0.631} & 0.783 \cellcolor[rgb]{0.720,0.906,0.670} & 0.713 \cellcolor[rgb]{0.644,0.879,0.714} & 1.000 \cellcolor[rgb]{0.997,0.953,0.572} & 0.996 \cellcolor[rgb]{0.992,0.952,0.568} & 0.995 \cellcolor[rgb]{0.992,0.952,0.568} \\
 &  & All Adv & 0.901 \cellcolor[rgb]{0.871,0.937,0.575} & 0.971 \cellcolor[rgb]{0.963,0.949,0.552} & 0.970 \cellcolor[rgb]{0.963,0.949,0.552} & 0.999 \cellcolor[rgb]{0.997,0.953,0.572} & 0.999 \cellcolor[rgb]{0.997,0.953,0.572} & 0.993 \cellcolor[rgb]{0.992,0.952,0.568} & 0.987 \cellcolor[rgb]{0.982,0.951,0.562} & 0.969 \cellcolor[rgb]{0.963,0.949,0.552} & 1.000 \cellcolor[rgb]{0.997,0.953,0.572} & 0.981 \cellcolor[rgb]{0.978,0.951,0.559} & 0.984 \cellcolor[rgb]{0.978,0.951,0.559} \\
\cline{2-11} 
 & \multirow{6}{*}{DFT} & No Adv* & 0.874 \cellcolor[rgb]{0.834,0.931,0.598} & 0.934 \cellcolor[rgb]{0.918,0.943,0.551} & 0.824 \cellcolor[rgb]{0.768,0.918,0.641} & 0.796 \cellcolor[rgb]{0.734,0.909,0.662} & 0.860 \cellcolor[rgb]{0.818,0.928,0.608} & 0.311 \cellcolor[rgb]{0.601,0.692,0.777} & 0.277 \cellcolor[rgb]{0.609,0.674,0.775} & 0.239 \cellcolor[rgb]{0.618,0.655,0.771} & 0.853 \cellcolor[rgb]{0.808,0.926,0.615} & 0.284 \cellcolor[rgb]{0.607,0.678,0.776} & 0.888 \cellcolor[rgb]{0.855,0.934,0.585} \\
 &  & No Adv & 0.882 \cellcolor[rgb]{0.844,0.933,0.591} & 0.971 \cellcolor[rgb]{0.963,0.949,0.552} & 0.801 \cellcolor[rgb]{0.744,0.912,0.656} & 0.808 \cellcolor[rgb]{0.748,0.913,0.653} & 0.914 \cellcolor[rgb]{0.886,0.939,0.566} & 0.235 \cellcolor[rgb]{0.619,0.653,0.771} & 0.236 \cellcolor[rgb]{0.619,0.653,0.771} & 0.218 \cellcolor[rgb]{0.623,0.642,0.768} & 0.971 \cellcolor[rgb]{0.963,0.949,0.552} & 0.165 \cellcolor[rgb]{0.634,0.612,0.756} & 0.910 \cellcolor[rgb]{0.886,0.939,0.566} \\
 &  & All GB-CO & 0.868 \cellcolor[rgb]{0.829,0.930,0.602} & 0.973 \cellcolor[rgb]{0.968,0.949,0.554} & 0.962 \cellcolor[rgb]{0.953,0.947,0.549} & 0.967 \cellcolor[rgb]{0.958,0.948,0.550} & 0.965 \cellcolor[rgb]{0.958,0.948,0.550} & 0.213 \cellcolor[rgb]{0.624,0.639,0.767} & 0.200 \cellcolor[rgb]{0.627,0.633,0.765} & 0.214 \cellcolor[rgb]{0.624,0.639,0.767} & 0.964 \cellcolor[rgb]{0.953,0.947,0.549} & 0.420 \cellcolor[rgb]{0.578,0.745,0.779} & 0.831 \cellcolor[rgb]{0.778,0.920,0.635} \\
 &  & All GB-DFT & 0.905 \cellcolor[rgb]{0.876,0.937,0.572} & 0.831 \cellcolor[rgb]{0.778,0.920,0.635} & 0.851 \cellcolor[rgb]{0.803,0.925,0.618} & 0.820 \cellcolor[rgb]{0.763,0.917,0.644} & 0.881 \cellcolor[rgb]{0.844,0.933,0.591} & 0.950 \cellcolor[rgb]{0.938,0.946,0.548} & 0.946 \cellcolor[rgb]{0.933,0.945,0.548} & 0.913 \cellcolor[rgb]{0.886,0.939,0.566} & 0.925 \cellcolor[rgb]{0.902,0.941,0.557} & 0.273 \cellcolor[rgb]{0.610,0.672,0.775} & 0.780 \cellcolor[rgb]{0.715,0.904,0.673} \\
 &  & All PGD & 0.959 \cellcolor[rgb]{0.948,0.947,0.548} & 0.930 \cellcolor[rgb]{0.912,0.942,0.553} & 0.809 \cellcolor[rgb]{0.753,0.914,0.650} & 0.784 \cellcolor[rgb]{0.720,0.906,0.670} & 0.875 \cellcolor[rgb]{0.834,0.931,0.598} & 0.114 \cellcolor[rgb]{0.640,0.580,0.736} & 0.089 \cellcolor[rgb]{0.642,0.563,0.722} & 0.089 \cellcolor[rgb]{0.642,0.563,0.722} & 0.943 \cellcolor[rgb]{0.928,0.944,0.549} & 0.992 \cellcolor[rgb]{0.987,0.952,0.565} & 0.935 \cellcolor[rgb]{0.918,0.943,0.551} \\
 &  & All Adv & 0.771 \cellcolor[rgb]{0.706,0.902,0.679} & 0.834 \cellcolor[rgb]{0.783,0.921,0.631} & 0.983 \cellcolor[rgb]{0.978,0.951,0.559} & 0.984 \cellcolor[rgb]{0.978,0.951,0.559} & 0.984 \cellcolor[rgb]{0.978,0.951,0.559} & 0.890 \cellcolor[rgb]{0.855,0.934,0.585} & 0.962 \cellcolor[rgb]{0.953,0.947,0.549} & 0.956 \cellcolor[rgb]{0.943,0.946,0.548} & 0.951 \cellcolor[rgb]{0.938,0.946,0.548} & 0.858 \cellcolor[rgb]{0.813,0.927,0.612} & 0.870 \cellcolor[rgb]{0.829,0.930,0.602} \\
\cline{2-11} 
 & \multirow{6}{*}{Direct} & No Adv* & 0.974 \cellcolor[rgb]{0.968,0.949,0.554} & 0.985 \cellcolor[rgb]{0.982,0.951,0.562} & 0.782 \cellcolor[rgb]{0.720,0.906,0.670} & 0.792 \cellcolor[rgb]{0.729,0.908,0.665} & 0.932 \cellcolor[rgb]{0.912,0.942,0.553} & 0.286 \cellcolor[rgb]{0.606,0.680,0.776} & 0.355 \cellcolor[rgb]{0.591,0.713,0.779} & 0.384 \cellcolor[rgb]{0.585,0.728,0.779} & 0.989 \cellcolor[rgb]{0.987,0.952,0.565} & 0.989 \cellcolor[rgb]{0.987,0.952,0.565} & 0.000 \cellcolor[rgb]{0.634,0.502,0.665} \\
 &  & No Adv & 0.947 \cellcolor[rgb]{0.933,0.945,0.548} & 0.989 \cellcolor[rgb]{0.987,0.952,0.565} & 0.868 \cellcolor[rgb]{0.829,0.930,0.602} & 0.862 \cellcolor[rgb]{0.818,0.928,0.608} & 0.965 \cellcolor[rgb]{0.958,0.948,0.550} & 0.279 \cellcolor[rgb]{0.608,0.676,0.775} & 0.369 \cellcolor[rgb]{0.588,0.721,0.779} & 0.427 \cellcolor[rgb]{0.577,0.748,0.779} & 0.992 \cellcolor[rgb]{0.987,0.952,0.565} & 0.991 \cellcolor[rgb]{0.987,0.952,0.565} & 0.013 \cellcolor[rgb]{0.636,0.510,0.674} \\
 &  & All GB-CO & 0.971 \cellcolor[rgb]{0.963,0.949,0.552} & 0.966 \cellcolor[rgb]{0.958,0.948,0.550} & 0.995 \cellcolor[rgb]{0.992,0.952,0.568} & 0.995 \cellcolor[rgb]{0.992,0.952,0.568} & 0.994 \cellcolor[rgb]{0.992,0.952,0.568} & 0.290 \cellcolor[rgb]{0.605,0.682,0.776} & 0.428 \cellcolor[rgb]{0.577,0.748,0.779} & 0.524 \cellcolor[rgb]{0.561,0.795,0.773} & 0.976 \cellcolor[rgb]{0.968,0.949,0.554} & 0.975 \cellcolor[rgb]{0.968,0.949,0.554} & 0.032 \cellcolor[rgb]{0.639,0.525,0.688} \\
 &  & All GB-DFT & 0.930 \cellcolor[rgb]{0.912,0.942,0.553} & 0.964 \cellcolor[rgb]{0.953,0.947,0.549} & 0.774 \cellcolor[rgb]{0.711,0.903,0.676} & 0.783 \cellcolor[rgb]{0.720,0.906,0.670} & 0.896 \cellcolor[rgb]{0.865,0.936,0.578} & 0.998 \cellcolor[rgb]{0.997,0.953,0.572} & 0.999 \cellcolor[rgb]{0.997,0.953,0.572} & 0.998 \cellcolor[rgb]{0.997,0.953,0.572} & 0.959 \cellcolor[rgb]{0.948,0.947,0.548} & 0.960 \cellcolor[rgb]{0.948,0.947,0.548} & 0.090 \cellcolor[rgb]{0.642,0.565,0.725} \\
 &  & All PGD & 0.979 \cellcolor[rgb]{0.973,0.950,0.556} & 0.976 \cellcolor[rgb]{0.968,0.949,0.554} & 0.849 \cellcolor[rgb]{0.803,0.925,0.618} & 0.848 \cellcolor[rgb]{0.803,0.925,0.618} & 0.963 \cellcolor[rgb]{0.953,0.947,0.549} & 0.193 \cellcolor[rgb]{0.629,0.628,0.763} & 0.287 \cellcolor[rgb]{0.606,0.680,0.776} & 0.363 \cellcolor[rgb]{0.590,0.717,0.779} & 0.991 \cellcolor[rgb]{0.987,0.952,0.565} & 0.988 \cellcolor[rgb]{0.982,0.951,0.562} & 0.997 \cellcolor[rgb]{0.997,0.953,0.572} \\
 &  & All Adv & 0.936 \cellcolor[rgb]{0.918,0.943,0.551} & 0.981 \cellcolor[rgb]{0.978,0.951,0.559} & 0.993 \cellcolor[rgb]{0.992,0.952,0.568} & 0.993 \cellcolor[rgb]{0.992,0.952,0.568} & 0.994 \cellcolor[rgb]{0.992,0.952,0.568} & 0.996 \cellcolor[rgb]{0.997,0.953,0.572} & 0.999 \cellcolor[rgb]{0.997,0.953,0.572} & 0.999 \cellcolor[rgb]{0.997,0.953,0.572} & 0.989 \cellcolor[rgb]{0.987,0.952,0.565} & 0.988 \cellcolor[rgb]{0.987,0.952,0.565} & 0.999 \cellcolor[rgb]{0.997,0.953,0.572} \\
\hline 
\multirow{9}{*}{MobileNet} & \multirow{3}{*}{Co-Occur} & No Adv* & 0.976 \cellcolor[rgb]{0.968,0.949,0.554} & 0.981 \cellcolor[rgb]{0.978,0.951,0.559} & 0.039 \cellcolor[rgb]{0.639,0.531,0.693} & 0.027 \cellcolor[rgb]{0.637,0.519,0.682} & 0.456 \cellcolor[rgb]{0.572,0.761,0.778} & 0.576 \cellcolor[rgb]{0.562,0.818,0.764} & 0.556 \cellcolor[rgb]{0.560,0.809,0.768} & 0.553 \cellcolor[rgb]{0.560,0.807,0.769} & 0.083 \cellcolor[rgb]{0.642,0.560,0.720} & 0.945 \cellcolor[rgb]{0.928,0.944,0.549} & 0.914 \cellcolor[rgb]{0.892,0.940,0.563} \\
 &  & No Adv & 0.974 \cellcolor[rgb]{0.968,0.949,0.554} & 0.985 \cellcolor[rgb]{0.982,0.951,0.562} & 0.029 \cellcolor[rgb]{0.638,0.522,0.685} & 0.020 \cellcolor[rgb]{0.637,0.516,0.679} & 0.359 \cellcolor[rgb]{0.590,0.715,0.779} & 0.695 \cellcolor[rgb]{0.626,0.871,0.724} & 0.672 \cellcolor[rgb]{0.610,0.863,0.733} & 0.647 \cellcolor[rgb]{0.590,0.851,0.744} & 0.036 \cellcolor[rgb]{0.639,0.528,0.691} & 0.947 \cellcolor[rgb]{0.933,0.945,0.548} & 0.924 \cellcolor[rgb]{0.902,0.941,0.557} \\
 &  & All Adv & 0.952 \cellcolor[rgb]{0.938,0.946,0.548} & 0.974 \cellcolor[rgb]{0.968,0.949,0.554} & 0.955 \cellcolor[rgb]{0.943,0.946,0.548} & 0.999 \cellcolor[rgb]{0.997,0.953,0.572} & 0.999 \cellcolor[rgb]{0.997,0.953,0.572} & 0.996 \cellcolor[rgb]{0.997,0.953,0.572} & 0.994 \cellcolor[rgb]{0.992,0.952,0.568} & 0.986 \cellcolor[rgb]{0.982,0.951,0.562} & 1.000 \cellcolor[rgb]{0.997,0.953,0.572} & 0.996 \cellcolor[rgb]{0.992,0.952,0.568} & 0.995 \cellcolor[rgb]{0.992,0.952,0.568} \\
\cline{2-11} 
 & \multirow{3}{*}{DFT} & No Adv* & 0.955 \cellcolor[rgb]{0.943,0.946,0.548} & 0.962 \cellcolor[rgb]{0.953,0.947,0.549} & 0.632 \cellcolor[rgb]{0.581,0.844,0.750} & 0.630 \cellcolor[rgb]{0.581,0.844,0.750} & 0.748 \cellcolor[rgb]{0.680,0.893,0.694} & 0.182 \cellcolor[rgb]{0.631,0.621,0.760} & 0.187 \cellcolor[rgb]{0.630,0.623,0.761} & 0.190 \cellcolor[rgb]{0.629,0.626,0.762} & 0.941 \cellcolor[rgb]{0.923,0.944,0.550} & 0.222 \cellcolor[rgb]{0.622,0.644,0.769} & 0.891 \cellcolor[rgb]{0.860,0.935,0.581} \\
 &  & No Adv & 0.928 \cellcolor[rgb]{0.907,0.942,0.555} & 0.970 \cellcolor[rgb]{0.963,0.949,0.552} & 0.524 \cellcolor[rgb]{0.561,0.795,0.773} & 0.539 \cellcolor[rgb]{0.560,0.802,0.771} & 0.770 \cellcolor[rgb]{0.706,0.902,0.679} & 0.192 \cellcolor[rgb]{0.629,0.628,0.763} & 0.189 \cellcolor[rgb]{0.629,0.626,0.762} & 0.176 \cellcolor[rgb]{0.632,0.619,0.759} & 0.926 \cellcolor[rgb]{0.907,0.942,0.555} & 0.217 \cellcolor[rgb]{0.623,0.642,0.768} & 0.910 \cellcolor[rgb]{0.881,0.938,0.569} \\
 &  & All Adv & 0.865 \cellcolor[rgb]{0.824,0.929,0.605} & 0.911 \cellcolor[rgb]{0.886,0.939,0.566} & 0.981 \cellcolor[rgb]{0.978,0.951,0.559} & 0.982 \cellcolor[rgb]{0.978,0.951,0.559} & 0.985 \cellcolor[rgb]{0.982,0.951,0.562} & 0.956 \cellcolor[rgb]{0.943,0.946,0.548} & 0.959 \cellcolor[rgb]{0.948,0.947,0.548} & 0.919 \cellcolor[rgb]{0.897,0.940,0.560} & 0.983 \cellcolor[rgb]{0.978,0.951,0.559} & 0.926 \cellcolor[rgb]{0.907,0.942,0.555} & 0.944 \cellcolor[rgb]{0.928,0.944,0.549} \\
\cline{2-11} 
 & \multirow{3}{*}{Direct} & No Adv* & 0.995 \cellcolor[rgb]{0.992,0.952,0.568} & 0.997 \cellcolor[rgb]{0.997,0.953,0.572} & 0.553 \cellcolor[rgb]{0.560,0.807,0.769} & 0.635 \cellcolor[rgb]{0.583,0.845,0.748} & 0.845 \cellcolor[rgb]{0.798,0.924,0.622} & 0.204 \cellcolor[rgb]{0.626,0.635,0.766} & 0.189 \cellcolor[rgb]{0.629,0.626,0.762} & 0.190 \cellcolor[rgb]{0.629,0.626,0.762} & 0.979 \cellcolor[rgb]{0.973,0.950,0.556} & 0.984 \cellcolor[rgb]{0.982,0.951,0.562} & 0.787 \cellcolor[rgb]{0.725,0.907,0.668} \\
 &  & No Adv & 0.993 \cellcolor[rgb]{0.992,0.952,0.568} & 0.996 \cellcolor[rgb]{0.992,0.952,0.568} & 0.640 \cellcolor[rgb]{0.585,0.847,0.747} & 0.694 \cellcolor[rgb]{0.626,0.871,0.724} & 0.808 \cellcolor[rgb]{0.748,0.913,0.653} & 0.088 \cellcolor[rgb]{0.642,0.563,0.722} & 0.043 \cellcolor[rgb]{0.640,0.534,0.696} & 0.030 \cellcolor[rgb]{0.638,0.522,0.685} & 0.849 \cellcolor[rgb]{0.803,0.925,0.618} & 0.861 \cellcolor[rgb]{0.818,0.928,0.608} & 0.744 \cellcolor[rgb]{0.676,0.892,0.696} \\
 &  & All Adv & 0.989 \cellcolor[rgb]{0.987,0.952,0.565} & 0.984 \cellcolor[rgb]{0.978,0.951,0.559} & 0.993 \cellcolor[rgb]{0.992,0.952,0.568} & 0.996 \cellcolor[rgb]{0.997,0.953,0.572} & 0.999 \cellcolor[rgb]{0.997,0.953,0.572} & 0.999 \cellcolor[rgb]{0.997,0.953,0.572} & 1.000 \cellcolor[rgb]{0.997,0.953,0.572} & 0.999 \cellcolor[rgb]{0.997,0.953,0.572} & 0.997 \cellcolor[rgb]{0.997,0.953,0.572} & 0.998 \cellcolor[rgb]{0.997,0.953,0.572} & 0.998 \cellcolor[rgb]{0.997,0.953,0.572} \\
\hline

\end{tabular}
}
\caption{Comprehensive table showing test set results on all datasets, for many training combinations. Each row represents a different detector and/or training set, and each column the data set tested on. The PGD data is generated using the models labeled with an asterisk.}
\label{tab:full_results_no_jpeg}
\end{table*}

All results are shown in table \ref{tab:full_results_no_jpeg}, with a summary of just the co-occurrence results in table \ref{tab:results_co_only}. These results show test accuracy on only one group at a time. When comparing across rows, it is important to note any changes in accuracy on real images in addition to changes in GAN performance.

From this table, we summarize the results as follows:

\begin{enumerate}
    \item Without adversarial retraining, all adversarial attacks would generally decrease performance on all detectors.
    \item Adversarially training on one attack method does not generally improve performance against other methods.
    \item After adversarial training, the models which are most different from the assumption in the adversarial attacks performed best. Notably, MobileNet trained on all adversarial images got over 98\% on all subsets.
\end{enumerate}

Of particular relevance to this paper is that for all co-occurrence based detectors which were not trained on the gray-box co-occurrence based attack, accuracy was less than 5\% for $\lambda=0$. This included models which were trained on all other adversarial attacks. This would seem to indicate that most co-occurrence detectors not trained on this particular attack would remain highly vulnerable. After retraining, accuracy on the $\lambda=0$ class was only slightly lower than regular GAN class. The difference was more significant in the MobileNet co-occurrence detector.

Also noteworthy is that the DFT and direct classifiers performed, on average, 18\% worse on the GB-CO $\lambda=0$ set than the original GAN images, if they were not trained against the GB-CO attack. Performance generally improved back to baseline levels after retraining.

\subsection{Other Tests}

\subsubsection{Reversed Gray-Box Co-Occurrence Attack}
\label{sec:reverse_co}

%While the concern with adversarial methods against GAN detectors is generally with fake images being passed as real, it is worth noting that the gray-box co-occurrence attack can also be used to trick a detector into classifying a real image as GAN. 

Though less useful as a real-world attack, the GB-CO method can also be used to generate real images which will be classified as GAN. With the target and source switched, we produced a test set of adversarial real images, and tested on the regular ResNet18 co-occurrence detector (row 1 in table \ref{tab:full_results_no_jpeg}). 95.9\% of the images were misclassified as GAN.

\subsubsection{Cross Channel Co-Occurrence Detector}
\label{sec:cband}

Recenty a paper was posted on arXiv from M. Barni \etal claiming to have improved the original co-occurrence GAN detector by including cross-channel co-occurrence matrices \cite{barni2020cnn}. Their cross-band co-occurrence matrices for a red-green pair are defined in equation \ref{eq:cband_co_occurrence}, assuming HWC convention on $X$. This is repeated for the red-blue and green-blue pairs. For spatial co-occurrence, they instead use diagonal pairs, shown in equation \ref{eq:diag_co_occurrence}. After producing these 6 co-occurrence matrices, they are stacked in the channel dimension, and passed to a ResNet18 classifier.

\begin{equation} \label{eq:cband_co_occurrence}
    C_{i,j} = \sum_{k,l} \delta(X_{k,l,1}-i) \cdot \delta(X_{k,l,2}-j)
\end{equation}

\begin{equation} \label{eq:diag_co_occurrence}
    C_{i,j} = \sum_{k,l} \delta(X_{k,l}-i) \cdot \delta(X_{k+1,l+1}-j)
\end{equation}

%We then trained a ResNet18 model on these new co-occurrence matrices produced on real and GAN images.
We also modified the co-occurrence gray-box attack equation in \ref{eq:diff_co_occurrence} to accept the 6 different pairs used in M. Barni \etal. We ran the algorithm to produce an adversarial test set for the cross-band co-occurrence detector. Results are shown in table \ref{tab:cband_results}. The gray-box attack still seems effective against detectors using this other co-occurrence feature.

\begin{table}
    \centering
    \scalebox{0.75}{
    \begin{tabular}{|c|c|c|}
        \hline
        Real & GAN & Adv C-Band \\
        \hline
        0.974 \cellcolor[rgb]{0.968,0.949,0.554} & 0.992 \cellcolor[rgb]{0.987,0.952,0.565} & 0.131 \cellcolor[rgb]{0.639,0.590,0.743} \\
        \hline
    \end{tabular}
    }
    \caption{Results of cross-channel co-occurrence detector on cross-channel co-occurrence method described in section \ref{sec:cband}, without adversarial retraining. Adversarial images were generated using the gray-box method, using this new co-occurrence formulation.}
    \label{tab:cband_results}
\end{table}

%cross band test: trained only on real and GAN images, but using cross-band

%test results
%0.9740276509961063 0.8812255859375 [[0.97858732 0.00750834 0.869     ]
% [0.02141268 0.99249166 0.131     ]]

%Reverse attack: real image as source and GAN image as taret

%0.05894145652381155 0.978515625 [[0.97886541 0.01640712 0.041     ]
% [0.02113459 0.98359288 0.959     ]]

\section{Conclusion}
In this paper, we presented two new attacks against co-occurrence based GAN detectors. We also demonstrate preliminary results showing that our attack can also generalize to at least one novel modification of our original co-occurrence function. Our work demonstrates the current vulnerability of three of the most popular detection methods to our adversarial attack, indicating the need for a real-world detector to be trained with these adversarial examples.

\ifwacvfinal
\section{Acknowledgements}
The real and GAN images used in this work were collected or generated from public sources during the first author's internship at Mayachitra Inc.

This work was partially supported by NSF SI2-SSI award \#1664172 and NSF HDR IDEAS2 award \#1934641. 
Any opinions, findings, and conclusions or recommendations expressed in this material are those of the author(s) and do not necessarily reflect the views of the National Science Foundation.

\fi

{\small
\bibliographystyle{ieee_fullname}
\bibliography{bib_files/detection_papers.bib,bib_files/gans_and_datasets.bib,bib_files/gen_adver.bib,bib_files/other.bib}
}

\end{document}